\documentclass[twocolumn,tighten,table]{aastex62}
\pdfoutput=1

\usepackage{amsmath,amstext}
\usepackage[T1]{fontenc}
\usepackage[figure, figure*]{hypcap}
\usepackage[tight,footnotesize]{subfigure}

\graphicspath{{./}{./figures/}}

\begin{document}
\title{Self-consistent Combined \emph{HST}, K-band, and Spitzer Photometric Catalogs of the BUFFALO Survey Fields}
\author[0000-0002-6015-8614]{Amanda Pagul}
\shortauthors{A. Pagul et al.}
\affiliation{Space Telescope Science Institute, 3700 San Martin Dr., Baltimore, MD 21218, USA}
\affiliation{Department of Physics and Astronomy, University of California Riverside, Pierce Hall, Riverside, CA 92521, USA}

\correspondingauthor{Amanda Pagul}
\email{apagul@stsci.edu}

\author[0000-0003-3136-9532]{F. Javier S\'{a}nchez}
\affiliation{Space Telescope Science Institute, 3700 San Martin Dr., Baltimore, MD 21218, USA}

\author[0000-0002-2951-7519]{Iary Davidzon}
\affiliation{Cosmic Dawn Center (DAWN)}
\affiliation{Niels Bohr Institute, University of Copenhagen, Lyngbyvej 2, Copenhagen \O~2100}

\author[0000-0002-6610-2048]{Anton M. Koekemoer}
\affiliation{Space Telescope Science Institute, 3700 San Martin Dr., Baltimore, MD 21218, USA}


\author{Hakim Atek}
\affiliation{Institut d'astrophysique de Paris, CNRS UMR7095, Sorbonne Universit\'e, 98bis Boulevard Arago, F-75014 Paris, France}

\author{Renyue Cen}
\affiliation{Department of Astrophysical Sciences, 4 Ivy Lane, Princeton, NJ 08544, USA}

\author[0000-0001-6278-032X]{Lukas J. Furtak}
\affiliation{Physics Department, Ben-Gurion University of the Negev, P. O. Box 653, Be'er-Sheva, 8410501, Israel}

\author[0000-0003-1974-8732]{Mathilde Jauzac}
\affiliation{Centre for Extragalactic Astronomy, Durham University, South Road, Durham DH1 3LE, U.K.}
\affiliation{Institute for Computational Cosmology, Durham University, South Road, Durham DH1 3LE, U.K}
\affiliation{Astrophysics and Cosmology Research Unit, School of Mathematical Sciences, University of KwaZulu-Natal, Durban 4041, South Africa}

\author[0000-0003-3266-2001]{Guillaume Mahler}
\affiliation{Centre for Extragalactic Astronomy, Durham University, South Road, Durham DH1 3LE, UK}
\affiliation{Institute for Computational Cosmology, Durham University, South Road, Durham DH1 3LE, UK}

\author{Bahram Mobasher}
\affiliation{Department of Physics and Astronomy, University of California Riverside, Pierce Hall, Riverside, CA 92521, USA}

\author[0000-0001-7847-0393]{Mireia Montes}
\affiliation{Instituto de Astrof\'{\i}sica de Canarias, c/ V\'{\i}a L\'actea s/n, E-38205 - La Laguna, Tenerife, Spain}
\affiliation{Departamento de Astrof\'isica, Universidad de La Laguna, E-38205 - La Laguna, Tenerife, Spain}   

\author[0000-0001-6342-9662]{Mario Nonino}
\affiliation{INAF-Trieste Astronomical Observatory}

\author[0000-0002-7559-0864]{Keren Sharon}
\affiliation{Department of Astronomy, University of Michigan, 1085 S. University Ave, Ann Arbor, MI 48109, USA}

\author[0000-0003-3780-6801]{Charles L. Steinhardt}
\affiliation{Cosmic Dawn Center (DAWN)}
\affiliation{Niels Bohr Institute, University of Copenhagen, Lyngbyvej 2, Copenhagen \O~2100}

\author[0000-0003-1614-196X]{John~R.~Weaver}
\affil{Department of Astronomy, University of Massachusetts, Amherst, MA 01003, USA}

\shorttitle{BUFFALO Catalogs}

\begin{abstract}
    
\end{abstract}
\begin{abstract}
This manuscript presents new astronomical source catalogs using data from the BUFFALO Survey. These catalogs contain detailed information for over 100,000 astronomical sources in the 6 BUFFALO clusters: Abell 370, Abell 2744, Abell S1063, MACS 0416, MACS 0717, and MACS 1149 spanning a total $240$ arcmin$^{2}$. The catalogs include positions and forced photometry measurements of these objects in the F275W, F336W, F435W, F606W, F814W, F105W, F125W, F140W, and F160W \textit{HST}-bands, Keck-NIRC2/VLT-HAWKI Ks band, and IRAC Channel 1 and 2 bands. Additionally, we include photometry measurements in the F475W, F625W, and F110W bands for Abell 370. This catalog also includes photometric redshift estimates computed via template fitting using \textsc{LePhare}. When comparing to spectroscopic reference, we obtain an outlier fraction of 9.2\% and scatter, normalized median absolute deviation (NMAD), of 0.062. The catalogs are publicly available for their use by the community.
\end{abstract}

\section{Introduction}
\label{sec:introduction}

\begin{figure*}
    \centering
    \includegraphics[width=0.32\textwidth]{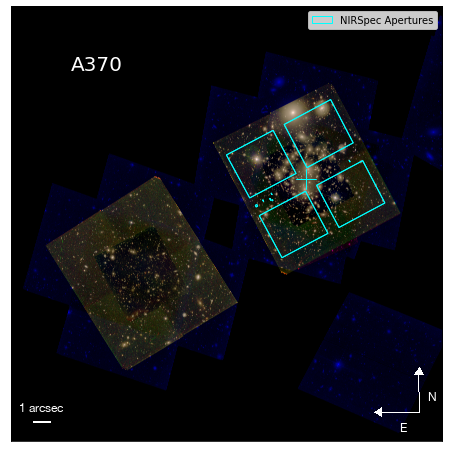}
   \includegraphics[width=0.32\textwidth]{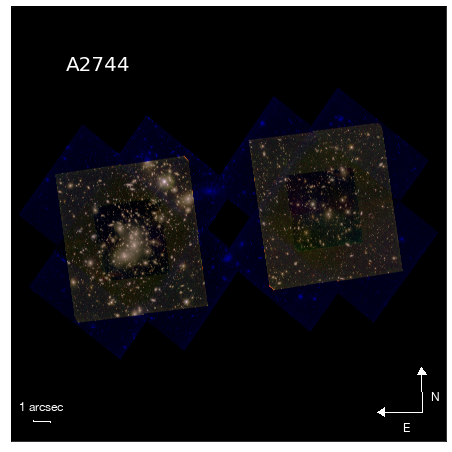}
    \includegraphics[width=0.32\textwidth]{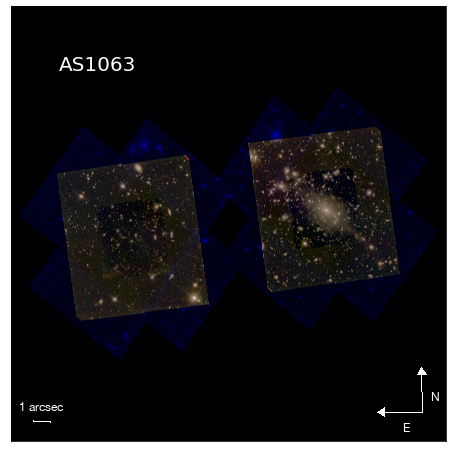}
    \includegraphics[width=0.32\textwidth]{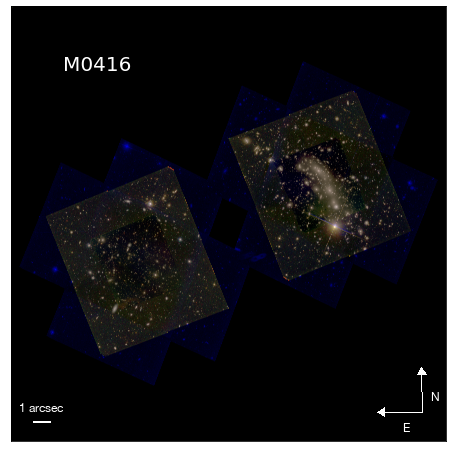}
    \includegraphics[width=0.32\textwidth]{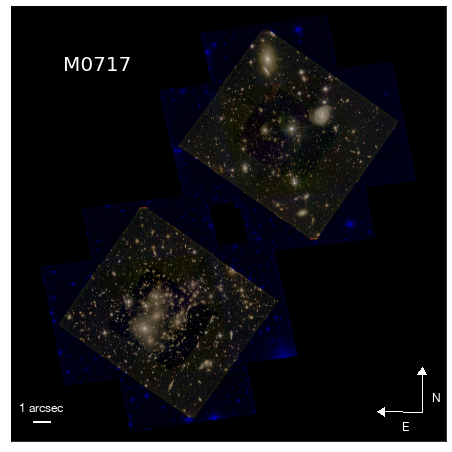}
    \includegraphics[width=0.32\textwidth]{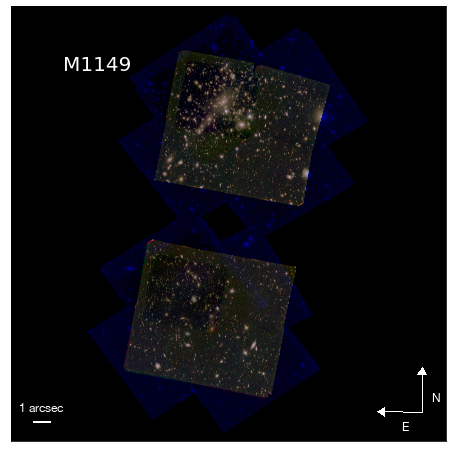}
    \caption{BUFFALO cluster footprints analyzed in this work. The HST mosaics were calibrated, aligned and created following the approaches described in \citet{koekemoer2011}, and the BUFFALO dataset is described in \citet{steinhardt20}. The RGB color pictures were created using \texttt{trilogy} \citep{2012ApJ...757...22C}, using F160W as the red channel, F105W as the green channel, and F814W as the blue channel. Note that due to the larger area coverage of F814W (ACS) compared to other bands (WFC3), certain areas of the footprint appear as blue. In the top left panel, we include the \emph{JWST} NIRSpec apertures for reference.}
    \label{fig:clusters}
\end{figure*}

The Hubble Frontier Fields (HFF) \citep{lotz17} is a multi-waveband program obtaining deep imaging observations of six massive clusters in a narrow redshift range $z\sim 0.308$ - $0.545$. Combining the sensitivity, resolution power and multi-wavelength capability of the Hubble Space Telescope (\textit{HST}), with the gravitational lensing effect introduced by the massive galaxy clusters selected for this study, one can reach unprecedented depths. Two \textit{HST} instruments, the Advanced Camera for Surveys (ACS) and Wide-Field Camera 3 (WFC3), were used in parallel to simultaneously observe each cluster and parallel field. The parallel fields separated by $\sim$ 6 arcmin from the cluster core, corresponding to $ >  1.8$ projected co-moving Mpc for a $z>0.3$ cluster. The six parallel fields are comparable in depth to the Hubble Ultra Deep Field \citep[HUDF,][]{2006AJ....132.1729B}, corresponding to m(AB) $\sim$ 29 mag. The area coverage and depth of the parallel fields provide significant improvement in the volume covered and statistics of faint galaxies. 

The aims of the HFF observations were: (1) leverage gravitational lensing due to massive clusters \citep[see][for a review]{kneib_natarajan11} to magnify fluxes and hence detect very faint background galaxies at $z$ $\sim 5$ - $10$ \citep[][and references therein]{schneider84,blandford_narayan86}. Strong lensing allows us to probe $\sim$ 2 magnitudes fainter than in blank fields. At the time of HFF observations, blank fields studies reached $\sim$-17 rest-frame UV magnitudes \citep{2015ApJ...810...71F,2015ApJ...803...34B}; (2) study the stellar population of these faint galaxies at high redshifts and constrain the mass function of galaxies at early epochs. Stellar masses reach down to $10^8 M_\odot$ in blank fields \citep[]{2016ApJ...825....5S,2021ApJ...922...29S,2022ApJS..258...11W,2022arXiv220711740K} and down to $10^6 M_\odot$ in HFF lensed fields \citep{2019MNRAS.486.3805B,2020ApJ...893...60K,2021MNRAS.501.1568F}; (3) study of the morphology and other observable properties of lensed galaxies at $z > 8$. 

The Beyond Ultra-deep Frontier Fields and Legacy Observations (\emph{BUFFALO}) is an \textit{HST} treasury program with 101 prime orbits (and 101 parallel orbits) (GO-15117; PIs: Steinhardt and Jauzac), covering the immediate areas around the HFF clusters where deep Spitzer (IRAC channels 1 and 2) and multi-waveband coverage already exist \citep{steinhardt20}. BUFFALO extends the spatial coverage of each of the six HFF clusters by three to four times. Observing these fields in five filters (ACS: F606W, F814W and WFC3: F105W, F125W and F160W), BUFFALO aims at a factor of 2 improvement in the statistics of high redshift galaxies \citep[][Pagul et al. in prep.]{2021MNRAS.501.1568F}, improves the cosmic variance and allows a more accurate modeling of the dark matter distribution in the foreground clusters. The \textit{HST} and Spitzer data for BUFFALO, combined with ground-based observations \citep[][KIFF]{KIFF} was specifically designed to expand the HFF to sufficiently large area to encompass a full James Webb Space Telescope NIRSpec field of view, without the need for \textit{JWST}/NIRCam pre-imaging. The program significantly improves the statistics of galaxies in the outskirts of clusters and field samples.

In this paper, we present photometric and redshift catalogs for the BUFFALO galaxies. The catalogs presented in this work aim to extend and complement previous efforts in the \emph{HFF} \citep{astrodeep1,astrodeep2,astrodeep3,astrodeep4,deepspace,nedkova21,2021ApJS..256...27P}. In section 2, we present the data used in this study. In section 3, we briefly outline the data reduction process, referring the reader to \citet{2021ApJS..256...27P} for a more detailed description. In section 4, we describe our photometric validation procedure. Section 5, details the data products and results. Section 6 describes the photometric redshifts extracted. Finally, our conclusions are presented in section 7.

Throughout this paper we assume standard cosmology with $\Omega_M = 0.23$, $\Omega_\Lambda = 0.76$ and $H_0 = 73$ Km/sec/Mpc. Magnitudes are in the AB system.

\section{The data}
\label{sec:dataset}

We provide a brief summary of the dataset in the following subsections. For more details about the design, aims and observations of BUFFALO we refer the reader to the BUFFALO overview paper \citep{steinhardt20}. All our data products are available at MAST as a High Level Science Product via \dataset[10.17909/t9-w6tj-wp63]{\doi{10.17909/t9-w6tj-wp63}}

\subsection{\textit{HST} observations}

The BUFFALO images provide the deepest exposures of galaxy clusters by \textit{HST}, only second to the HUDF with respect to depth. With 101 additional prime (and 101 parallel) orbits, they build on the existing HFF cluster and parallel field surveys. BUFFALO slightly increases the depth at the center of the HFF clusters while increasing their areal coverage three- to four fold. As a result, it expands the radial coverage of cluster outskirts, providing observations of the global mass distribution of clusters to almost the virial radius, i.e. $\sim 3/4 \times R_{vir}$. The coverage was chosen to increase the high-z sample size, in particular for rare bright high-mass galaxies at $z\sim8-9$. Furthermore, BUFFALO's footprint is chosen to be compatible with \textit{JWST}'s NIRSpec field of view, allowing multiwavelength programs with \textit{JWST}\footnote{These were produced using the \texttt{JWST\_footprints} module (\url{https://github.com/spacetelescope/JWST\_footprints}).} (Figure \ref{fig:clusters}), which is especially timely for planning robust observations with \textit{JWST}.

In the HFF, the gravitational potential of the clusters' halo, besides binding together the galaxies in the system, produces a lensing magnification that could detect background objects to apparent magnitudes of 30--33 mag, i.e.\ 10--100 times fainter than previous surveys. With BUFFALO, we get magnifications of $\sim 4$ on average. Details of the BUFFALO survey design are provided in \citet{steinhardt20}.  In Table \ref{tab:surveyprop}, we report the main characteristics of the six clusters, with a summary of the ancillary observations in Table~\ref{tab:otherdata}. We use the official BUFFALO mosaics, with a pixel scale of 0.06"/pix, which have been produced following the procedures outlined in \citet{koekemoer2011}; the full BUFFALO dataset is described further in \citet{steinhardt20}. 

We complement this data with the available public F275W and F336W \emph{HFF} data from the HFF-Deepspace campaign \citep{deepspace}, which uses observations from \citet{alavi16}.

\subsection{Ancillary data}

The large wealth of complementary legacy datasets and programs for the HFF clusters has contributed to its success. The \textit{Spitzer} Space Telescope dedicated more than 1,000 hours of Director's Discretionary time to obtain Infrared Array Camera (IRAC) 3.6\,$\mu$m (channel 1) and 4.6\,$\mu$m (channel 2) imaging down to the depths of 26.5 and  26.0\,mag., in cluster and parallel fields respectively (program IDs: Abell 2744: 83, 90275; MACS J0416.1-2403: 80168, 90258; MACS J0717.4+3745: 40652, 60034, 90009, 90259; MACS J1149.4+2223: 60034, 90009, 90260; Abell S1063 (RXC J2248.7-4431): 83, 10170, 60034; Abell 370: 137, 10171, 60034). These observations are especially important for redshift determination given that they help break the degeneracies between low-redshift interlopers and high-redshift galaxies, and are beneficial in constraining galaxy properties since they provide a good proxy for galaxy stellar mass.




The HFF clusters in the southern sky are also covered in the \textit{Ks} band using the High Acuity Wide Field K-band Imager (HAWK-I) \citep[KIFF][]{KIFF,pirard04,2008A&A...491..941K} at the Very Large Telescope (VLT), reaching a depth of 26.0\,mag (5$\sigma$, point-like sources) for Abell 2744, MACS-0416, Abell S1063, and Abell 370 clusters. In the northern sky, this campaign used the Multi-Object Spectrometer for Infrared Exploration (MOSFIRE) \citep{2010SPIE.7735E..1EM,2012SPIE.8446E..0JM} at Keck to observe MACS-0717 and MACS-1149 to a K-band 5$\sigma$ depth of 25.5 and 25.1\,mag respectively. This data covers all of the cluster and parallel field centers, but not the entirety of the outer area observed by BUFFALO. Table \ref{tab:otherdata} summarizes the available ancillary data.

\begin{deluxetable*}{lcccccc}
\label{tab:surveyprop}
\tablecaption{
\label{tab:surveyprop}
	Frontier Field cluster and parallel field positions, along with clusters' mean redshift ($z_\mathrm{clu}$), virial mass ($M_\mathrm{vir}$), and X-ray luminosity ($L_X$) \citep{lotz17}}

\tablehead{%
Field	        	& Cluster Center (J2000) &	Parallel Center (J2000)	& $z_\mathrm{clu}$  & $M_{vir}$ & $L_{X}$ \\ 	        	& R.A., Decl. &	R.A., Decl.	& &  &  }
\startdata
Abell 370           &  02:39:52.9, -01:34:36.5 & 02:40:13.4, -01:37:32.8 & 0.375 & $\sim 1\times10^{15}$ & $1.1\times10^{45}$\\
					&		    &   	&   	        	\\
Abell 2744		    &  00:14:21.2, -30:23:50.1 & 00:13:53.6, -30:22:54.3  & 0.308  & $1.8 \times 10^{15}$  &  $3.1\times10^{45}$ \\
        			&	    	&   	&   			        	\\
Abell S1063		    & 22:48:44.4, -44:31:48.5 & 22:49:17.7, -44:32:43.8  & 0.348 & $1.4\times10^{15}$ & $1.8\times10^{45}$\\
   &		    &   	&   	        	\\
MACS J0416.1-2403	&  04:16:08.9, -24:04:28.7 & 04:16:33.1, -24:06:48.7   & 0.396 & $1.2 \times 10^{15}$ &  $1.0\times10^{45}$   &\\       	      	
					&	    	&   &	&     \\
MACS J0717.5+3745   &  07:17:34.0 +37:44:49.0 & 07:17:17.0 +37:49:47.3 & 0.545 & $\sim 2-3\times10^{15}$ & $3.3\times10^{45}$\\
					&		    &   	&   	        	\\
MACS J1149.5+2223	&  11:49:36.3, +22:23:58.1 & 11:49:40.5, +22:18:02.3   & 0.543 & $2.5\times10^{15}$  &  $1.8\times10^{45}$	\\
	& & & & 
\enddata
\end{deluxetable*}

\begin{deluxetable*}{lccccc}
\tablecaption{\label{tab:otherdata}
	Existing multi-wavelength HFF coverage from follow-up programs, as used in the present work. The 5-$\sigma$ point-source depth was estimated by integrating the noise in a 2D Gaussian PSF aperture with the FWHM value set to the ones given in Table \ref{tab:psffwhm}. The HFF \citep{lotz17} program is led by PIs T. Soifer and P. Capak; KIFF PI is G. Brammer \citep{KIFF}.}

\tablehead{%
Field	        	& Observatory/Camera	& Central Wavelength	& Depth}
\startdata
Abell 370           & VLT/HAWK-I   & 2.2$\mu m$  & $\sim$ 26.18 \\
          & \textit{Spitzer} IRAC 1,2  & 3.6$\mu m$, 4.5$\mu m$ & $\sim$ 25.19, 25.09 	\\
					&		    &   	&   	        \\
					MACS J0717.5+3745	& Keck/MOSFIRE   & 2.2$\mu m$  & $\sim$ 25.31  \\
	& \textit{Spitzer} IRAC 1,2   & 3.5$\mu m$, 4.5 $\mu m$  &  $\sim$ 25.04, 25.17  \\
					&		    &   	&   	        	\\
MACS J0416.1-2403	& VLT/HAWK-I   & 2.2$\mu m$  & $\sim$ 26.25  \\
	& \textit{Spitzer} IRAC 1,2  & 3.5$\mu m$, 4.5 $\mu m$  & $\sim$ 25.31, 25.44    \\        	      	
					&	    	&   &	&     \\
Abell S1063		    & VLT/HAWK-I   & 2.2$\mu m$  & $\sim$ 26.31   \\
   & \textit{Spitzer} IRAC 1,2  & 3.6$\mu m$, 4.5$\mu m$ & $\sim$ 25.04, 25.04     \\
   &		    &   	&   	        	\\
Abell 2744		    & VLT/HAWK-I   & 2.2$\mu m$  & $\sim$ 26.28  \\
	    & \textit{Spitzer} IRAC 1,2  & 3.6$\mu m$, 4.5$\mu m$ & $\sim$ 25.32, 25.08  \\
        			&	    	&   	&   			        	\\
MACS J1149.5+2223	& Keck/MOSFIRE   & 2.2$\mu m$  & $\sim$ 25.41  \\
	& \textit{Spitzer} IRAC 1,2   &  3.5$\mu m$, 4.5 $\mu m$ &  $\sim$ 25.24, 25.01 \\
	& & & & 
\enddata
\end{deluxetable*}






\begin{deluxetable}{lccc}[t]
\tablecaption{The Point Spread Function radius and effective wavelengths for different photometric bands used for the BUFFALO fields.}
\label{tab:psffwhm}
\tablehead{\colhead{Band} & \colhead{\hspace{.75cm}$FWHM$}\hspace{.5cm} & \colhead{$\lambda_{\mathrm{pivot}}$ (\AA)}}
\startdata
F275W & 0\farcs11 & 2710\\
F336W & 0\farcs12 & 3354\\
F435W & 0\farcs13 & 4329\\
F606W & 0\farcs11 & 5922\\
F814W & 0\farcs10 & 8045\\
F105W & 0\farcs20 & 10551\\
F125W & 0\farcs20 & 12486\\
F140W & 0\farcs20 & 13923\\
F160W & 0\farcs20 & 15369\\
Ks & 0\farcs36 & 21524\\
I1 & 1\farcs29 & 35634\\
I2 & 1\farcs42 & 45110\\
\enddata
\tablecomments{Values were calculated for the cluster Abell 370.}
\end{deluxetable}


\section{Data processing}

The workflow followed for the data processing in this work is the same as the one in~\citet{2021ApJS..256...27P} (P21 hereafter). The main steps taken to obtain the data products presented here are summarized as follows:
\begin{enumerate}
    \item Error map correction: we compare the standard deviation of the values of the background pixels in the science image, with the reported root mean-square (rms) values as given by the error maps, and correct the latter so that the mean ratio in the background pixels are equal to $1$.
    \item PSF extraction: we select unsaturated, unblended stars and perform median stacking to obtain an estimate of the PSF.
    \item Intracluster light (ICL) + bright galaxy modeling: Perform multi-object fits to S\'{e}rsic profiles, plus a local background using a combination of \texttt{GALFIT}~\citep{2010AJ....139.2097P} and \texttt{GALAPAGOS-M}~\citep{2013MNRAS.430..330H}.
    \item Bright galaxy photometry: we run \textsc{Source Extractor}~\citep{1996A&AS..117..393B} on \textit{HST} bands PSF-matched to the reddest, F160W, band, and obtain photometric measurements.
    \item Background galaxy photometry: we subtract the bright galaxies and ICL, and run \textsc{Source Extractor} on the ``cleaned'' field for the PSF-matched \textit{HST} images.
    \item Spitzer and K-band photometry: we use T-PHOT~\citep{2016A&A...595A..97M} to obtain self-consistent photometry measurements on the Spitzer and K-band images, using the \textit{HST} images and segmentation maps as priors.
    \item Synthetic source injection: we inject synthetic sources and repeat the process to validate and correct the photometric measurements.
    \item Estimate photometric redshifts: the last step consists on using \textsc{LePhare}~\citep{LePhare1999, LePhare2006} to obtain photometric redshift estimates of detected galaxies in these catalogs.
\end{enumerate}
In the following subsections some of these steps are described in more detail. For a detailed description of all the steps, we refer the reader to P21.

\subsection{Point Spread Function}
\label{ssec:psf}
\begin{figure*}
    \centering
    \includegraphics[width=\textwidth]{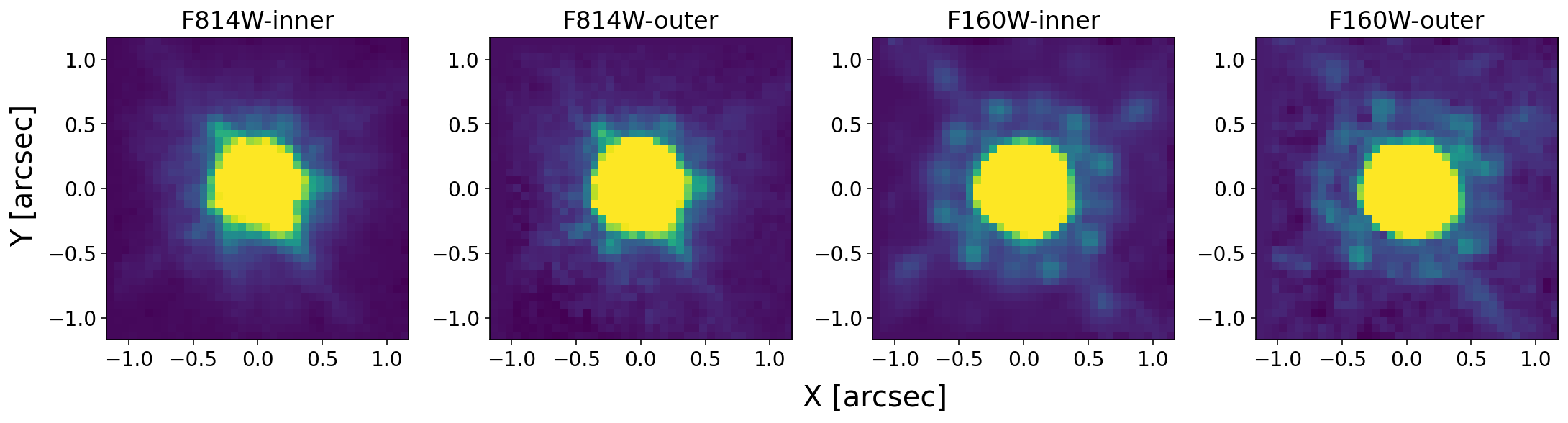}
    \caption{Representative examples of the point spread function (PSF) for the instruments used in this study, corresponding to a 0.06"/pixel scale normalized with the ZScale algorithm. From left to right, panels show ACS-F814W (inner region), ACS-F814W (outer region), WFC3-F160W (inner region), WFC3-F160W (outer region). See Section 3.1 for more details}
    \label{fig:psfs_inout}
\end{figure*}

A well-defined point spread function (PSF) as a function of wavelength is crucial to perform consistent photometry within a `panchromatic' baseline to correctly model galaxies and obtain galaxy fluxes in PSF-matched images. In order to perform multi-waveband photometry with accurate signal-to-noise and resolution for each aperture, we convolve images with a kernel generated by taking (in Fourier space) the ratio between their original and target PSFs, to match that of the reddest F160W PSF. In order to generate the PSFs for the \textit{HST} and K-band images, we stack isolated and unsaturated stars in each individual image, taking the median of the stack. Up to this point, the procedure is identical to that followed in P21. We improve upon our previous work by creating PSFs for the representative inner (deeper) and outer (shallower) regions in both the cluster- and parallel-fields. Figure \ref{fig:psfs_inout} shows examples of the stacked PSFs derived in different regions, and Table \ref{tab:psffwhm} gives the representative FWHM as a function of wavelength. We note that the full-width-half-max (FWHM) in both regions are compatible. 

Due to large spatial variations of the PSF in the mid-IR \emph{Spitzer} channels \footnote{
See \href{https://irsa.ipac.caltech.edu/data/Spitzer/docs/irac/calibrationfiles/psfprf/}{the \emph{Spitzer}/IRAC handbook}}, we do not use the same approach to create our \emph{Spitzer} PSF model. Furthermore, the individual pixel response functions (PRFs) are asymmetric and are thus dependent on the orientation of the camera.  Moreover, the pixels on IRAC Ch 1 and 2 tend to under sample the PRF\footnote{More information in the  \href{https://irsa.ipac.caltech.edu/data/Spitzer/docs/files/Spitzer/simfitreport52_final.pdf}{the \emph{Spitzer}/IRAC handbook}.}. Thus, instead of stacking stars and generating a single PSF per field, we use a synthetic pixel response function (PRF) that combines the information on the PSF, the detector sampling, and the intrapixel sensitivity variation in response to a point-like source, as done in P21. A PRF model for a given position on the IRAC mosaic is generated by the code \texttt{PRFMap} (A.\ Faisst, private communication)  by combining the single-epoch frames that contribute to that mosaic. To do so, \texttt{PRFMap} stacks individual PRF models  with the same orientation of the frames, resulting in a realistic, spatially-dependent PSF model.

\subsection{Modeling the intra-cluster light}
The deep potential well and high density of galaxy clusters make them rich laboratories to study galaxy dynamics and interactions. Due to these complex processes, stars and gas stripped from their constituent galaxies build up in the cluster core as intracluster light (ICL) \citep[see][for a review]{2022NatAs...6..308M}. This can bias the flux measurements of galaxies, close in angular space, to the cluster center. Following~\citet{2017ApJ...846..139M} and P21, in order to model the ICL in the BUFFALO clusters, we first generate $18\times18$ arcsecond ($300\times300$ pixel) stamps centered on each galaxy with a magnitude brighter than 26 in each image/band. Using \texttt{GALFIT}~\citep{2010AJ....139.2097P}, we fit all galaxies in each stamp with a single S\'{e}rsic profile, masking those that are fainter than magnitude 26. In case a given pixel with coordinates $(x, y)$ is only included in one cutout, the ICL emission ($F_\mathrm{ICL}$) is defined as the local background measurement as reported by \texttt{GALFIT} (namely, the \texttt{sky value} parameter). If there are overlapping cutouts in $(x, y)$, we use the inverse $\chi^2$-weighted mean of their background measurements:
\begin{eqnarray}
\label{eq:FICL}
F_\mathrm{ICL}(x,y)=\frac{\Sigma_{i} s_i(x,y)/\chi^2_i(x,y)}{\Sigma_{i} 1/\chi^2_i(x,y)} \;,
\end{eqnarray}
where $s_{i}$ and $\chi_i^2$ are the \texttt{sky fit} (fit value to the local background of the postage stamp)  and goodness-of-fit values from \texttt{GALFIT} for the $i$-th cutout, respectively.

As described in P21, the resulting ICL map has unphysical sharp features, which are smoothed out using a Gaussian kernel with $\sigma=4.32"$.

Similarly, for the $K_s$ and $Spitzer$ bands, we use \texttt{T-PHOT} to obtain the local background for each measured source, which is then merged into a single mosaic, and smoothed with a representative kernel.

As a caveat, though these maps primarily contain ICL emission, they also contain inhomogeneities in the background. This ensures a robust `background$+$ICL subtraction" in the individual images. Cleaning of these maps via color selection of the individual stamps will then be performed.

\subsection{Modeling the brightest galaxies}
\label{ssec:bcg_icl_model}
\begin{figure*}
    \centering
    \includegraphics[width=\textwidth]{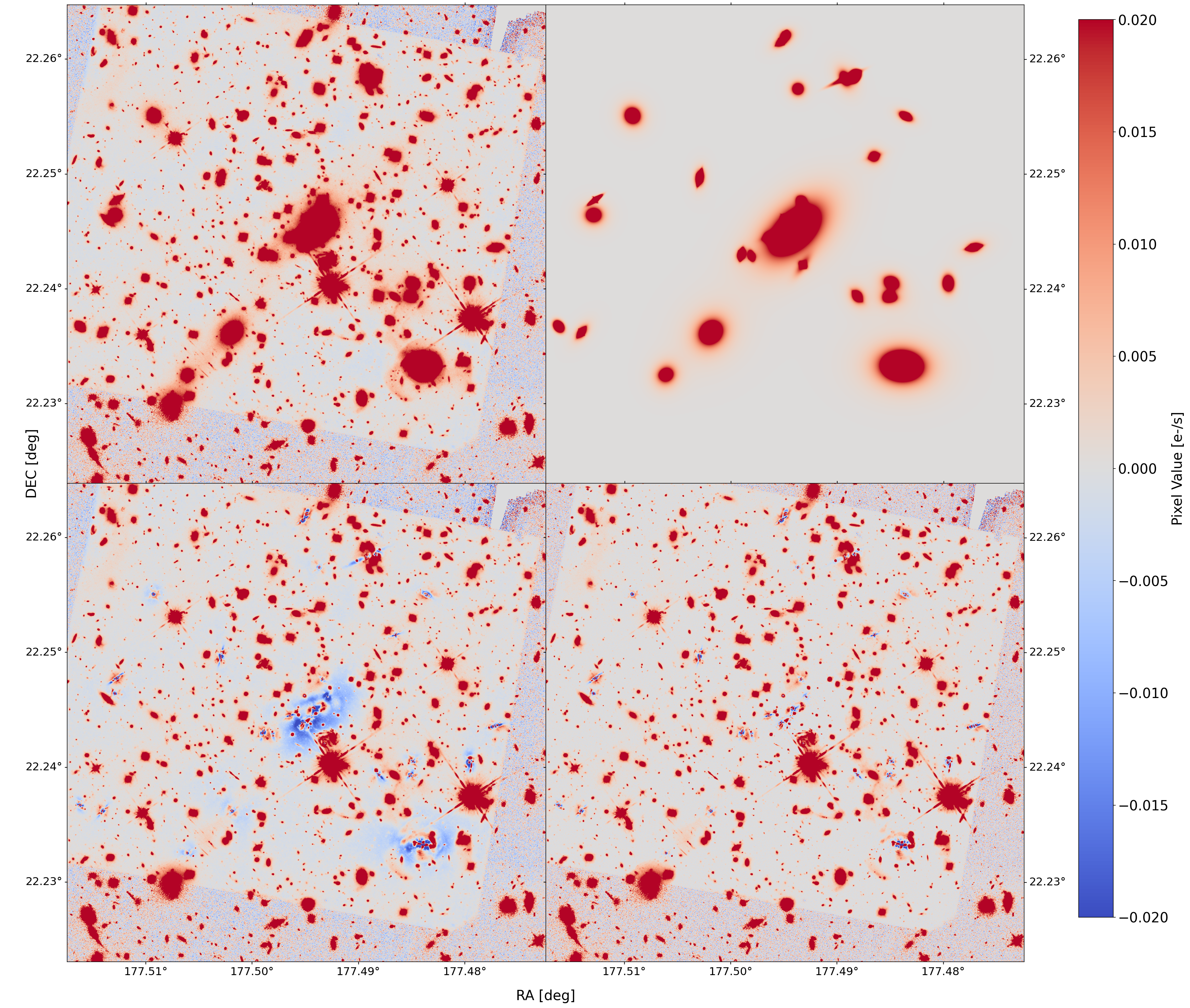}
    \caption{Steps in bright cluster + ICL modelling (in this case for cluster MACS 1149) for the F160W band. Upper panels show the original image (left) and the galaxy/ICL models (right). Lower panels show the residual image before and after median filtering (left- and right-hand panel respectively). The colorbar denotes the pixel intensity in counts/s. See Sections from 3.2 to 3.3 for more details.}
    \label{fig:fov}
\end{figure*}

The procedure to model bright galaxies (magnitude brighter than 19) is also unchanged from P21. We rely on GALAPAGOS-M~\citep{2013MNRAS.430..330H} to fit S\'{e}rsic profiles simultaneously to galaxies in all bands, with the fitting parameters varying as a function of wavelength. We construct galaxy models for the relevant galaxies and also cross-check the fits with those in \citet{nedkova21}. The results of the ICL and bright galaxy modeling and subtraction are illustrated in Figure~\ref{fig:fov}. 

Finally, we apply a median filter to the ICL+bright galaxy subtracted images. We use a filter with a box size of 1$^\circ$ per side, applied only to pixels within 1$\sigma$ of the background level to reduce the effects of over-subtraction in the residual. Figure \ref{fig:fov} shows the modeling and filtering process. The lower right panel shows the effect of median filtering. Note that this process does not significantly affect the
outskirts of the cluster.

\begin{figure}
\includegraphics[width=0.4\textwidth]{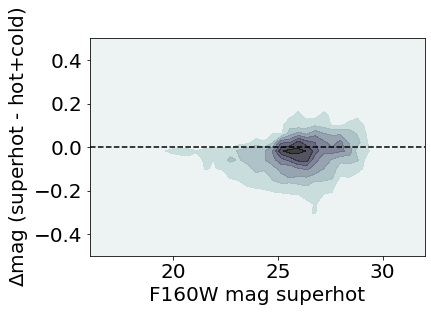}
\caption{Magnitude comparison between ``hot+cold" mode as described in P21, and the ``super hot" mode used in this work. Note, the magnitude difference is primarily within 0.05 mags.}
\label{fig:coldhot_v_superhot}
\end{figure}

\subsection{Source Extraction}
To detect galaxies and perform photometry, we use \textsc{Source Extractor}, focusing only on the "super hot" mode, rather than creating a dual run with hot and cold modes (see P21 for definition of "hot" and "cold" modes). This is one of the main differences with the procedure presented in P21 where a second ``cold" mode \texttt{Source Extractor} run is performed. We find that this second run does not have a significant impact on the detection nor photometric performance ($< 0.05$ mag), especially after bright galaxy and ICL subtraction. This is a consequence of the cold mode focusing on extracting information about the brightest objects, which have already been removed by the bright galaxy subtraction. This is illustrated in Figure~\ref{fig:coldhot_v_superhot}, where we compare a dual run with our new ``super hot" run, finding similar magnitudes for the BUFFALO cluster Abell 370. The final \textsc{Source Extractor} configuration file is presented in Appendix \ref{app:Source Extractor_config}.

\begin{figure*}
    \centering
    \includegraphics[width=0.95\textwidth]{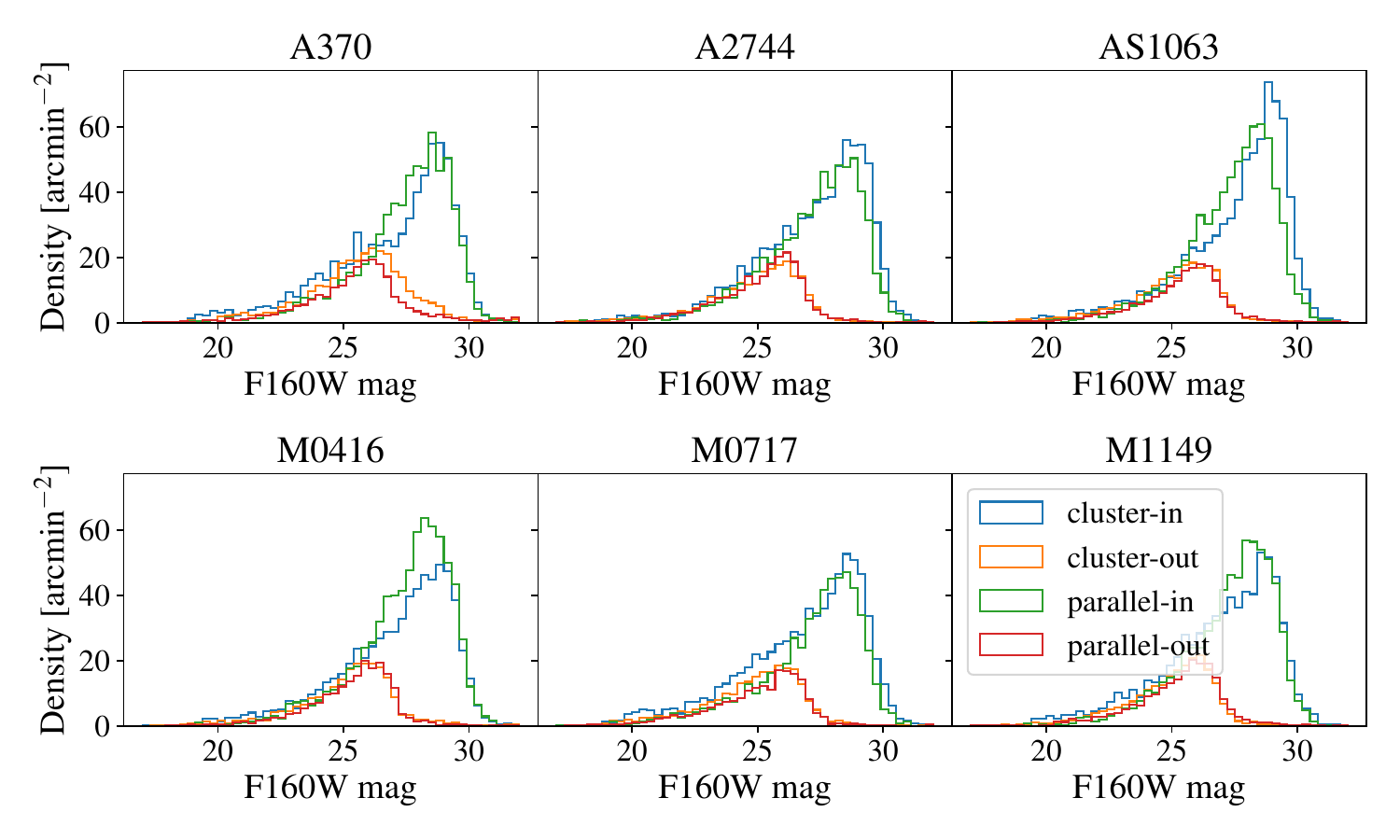}
    \caption{Magnitude distributions for sources in the BUFFALO catalogs across all clusters. We subdivide each of the catalogs, one per cluster and per parallel/infall field, into inner (in) and outer (out) regions, which correspond to different depth regimes.}
    \label{fig:mag_dists}
\end{figure*}

We also show the magnitude distribution of sources in the F160W band for all clusters in Figure~\ref{fig:mag_dists}. The large number density (defined as the number of sources per square arcmin) and depth of these catalogs are indicated. We subdivided the catalogs into sources detected in the inner field regions (the overlap with HFF), which reaches to significant depth, and the outer regions (the extension), where the depth is noticeably lower. The differences between the distributions of the cluster and the parallel regions is apparent. The cluster regions typically contain an over-abundance of brighter galaxies, whereas the parallel fields contain less of these bright objects but reach slightly deeper levels.

\subsection{Photometry in Ancillary images}
Because the \textit{Ks} and \textit{Spitzer} images have lower angular resolution than the \textit{HST} images, they are more affected by blending. In order to effectively deblend sources and maximize the information extracted in each image, we use \textsc{T-PHOT} as in P21 to perform forced photometry in the Ks- and IRAC images on sources detected in the IR-Weighted \textit{HST} image. \textsc{T-PHOT} \citep{2015A&A...582A..15M,2016A&A...595A..97M} is a software that uses priors from high resolution data in order to deblend and extract fluxes of the same objects in a lower resolution image. We first use \textsc{T-PHOT}'s built-in background routine to generate a local background for each source and remove the excess ICL light as well as inhomogenieties in the backgrounds. Then, as ``real'' galaxy priors, we use the IR-Weighted segmentation map and flux measurements from the F160W-band image. Additionally, we use the galaxy models that have been created in the bright galaxy$+$ICL removal step as the ``model'' priors. Given the spatial variation of the PRF in the IRAC bands, we take advantage of \textsc{T-PHOT}'s ``multikernel'' option, and use a separate PRF to model sources at each position. We emphaize that the flux (FitQty) that is provided by \textsc{T-PHOT} corresponds to the total flux emitted by a given source.


\section{Photometric validation}
\label{sec:sims}

In order to characterize the performance of our detection and measurement procedures, we proceed as in P21 injecting synthetic galaxies in the original BUFFALO images using \texttt{GalSim}~\citep{2015A&C....10..121R} to render noiseless realistic galaxies via the \texttt{RealGalaxy} class following the morphology measurements in \textit{COSMOS} by~\citet{2007ApJS..172..219L}. This catalog only contains information for fluxes in the F814W band. Thus, we match these sources to the COSMOS catalog~\citep{2016ApJS..224...24L} in order to obtain the fluxes in the rest of our bands of interest. We choose to keep the morphology and centroids fixed across bands in order to simplify data handling and bookkeeping. In this case, we generate 10 realizations of a set of 160 sources using the F160W image footprint as reference. Note that, since not all bands cover the same footprint, some sources will not be recovered after processing. We then insert these sources in the original images, run our pipeline on the resulting combined image (which is the sum of the original and the noiseless synthetic sources) and compare their measured fluxes and positions to their inputs. 

\begin{figure*}
    \centering
    \includegraphics[width=0.3\textwidth]{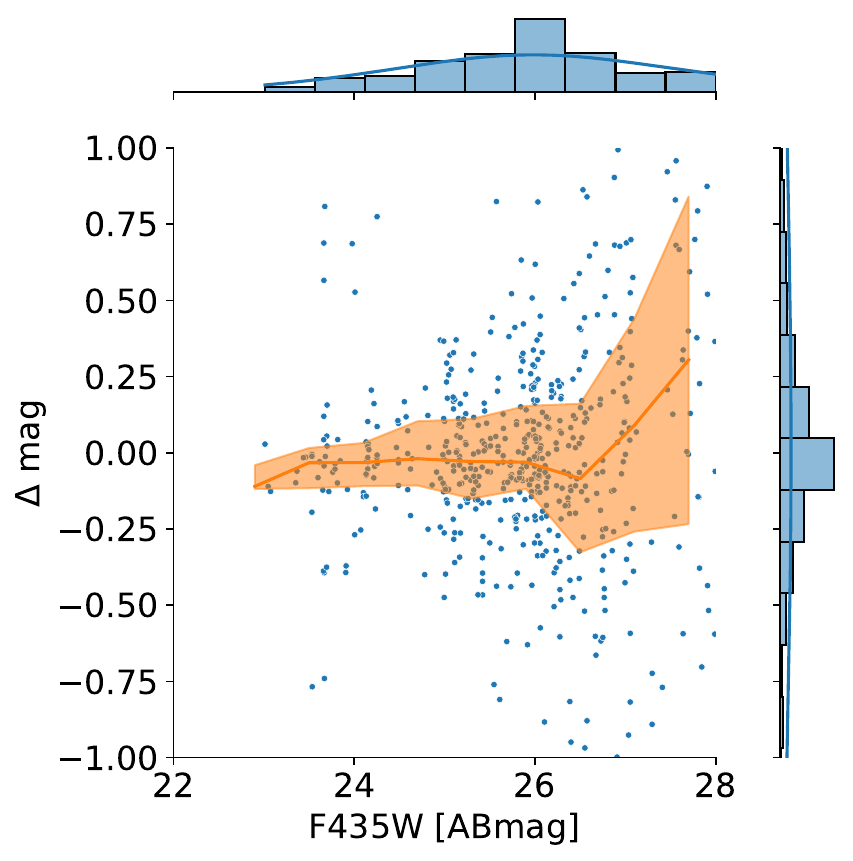}
    \includegraphics[width=0.3\textwidth]{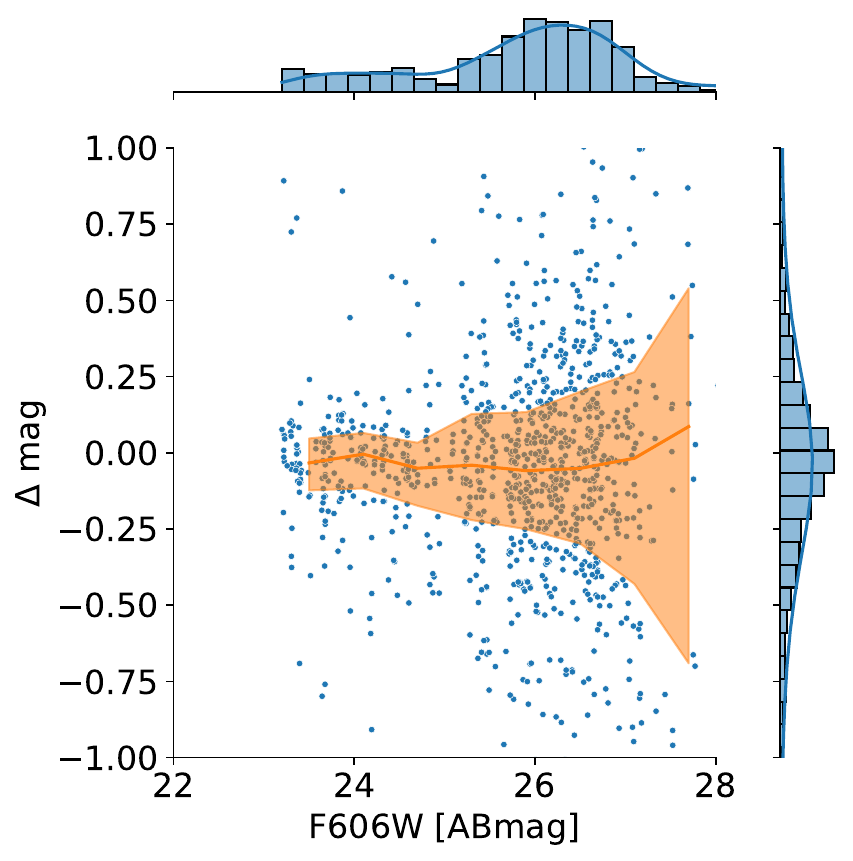}
    \includegraphics[width=0.3\textwidth]{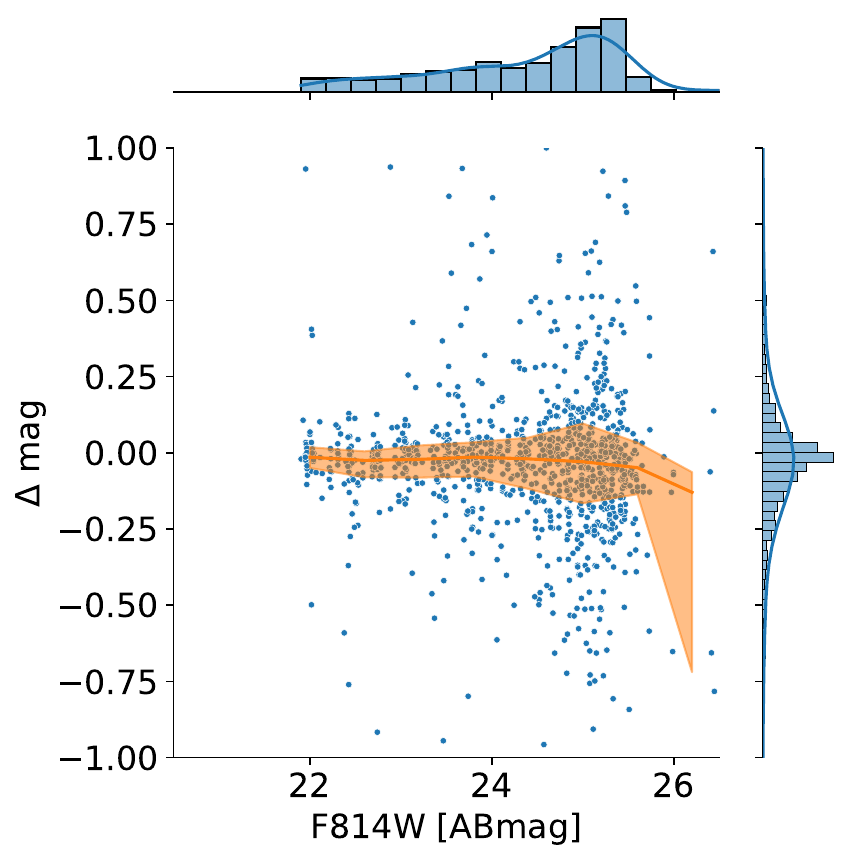}
    \includegraphics[width=0.3\textwidth]{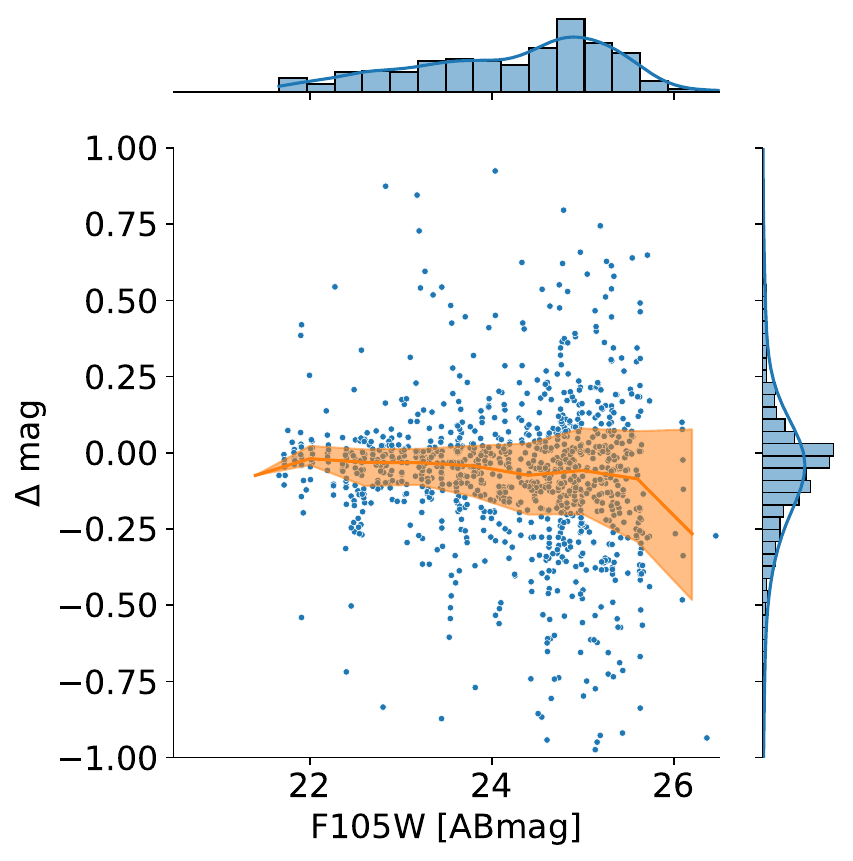}
    \includegraphics[width=0.3\textwidth]{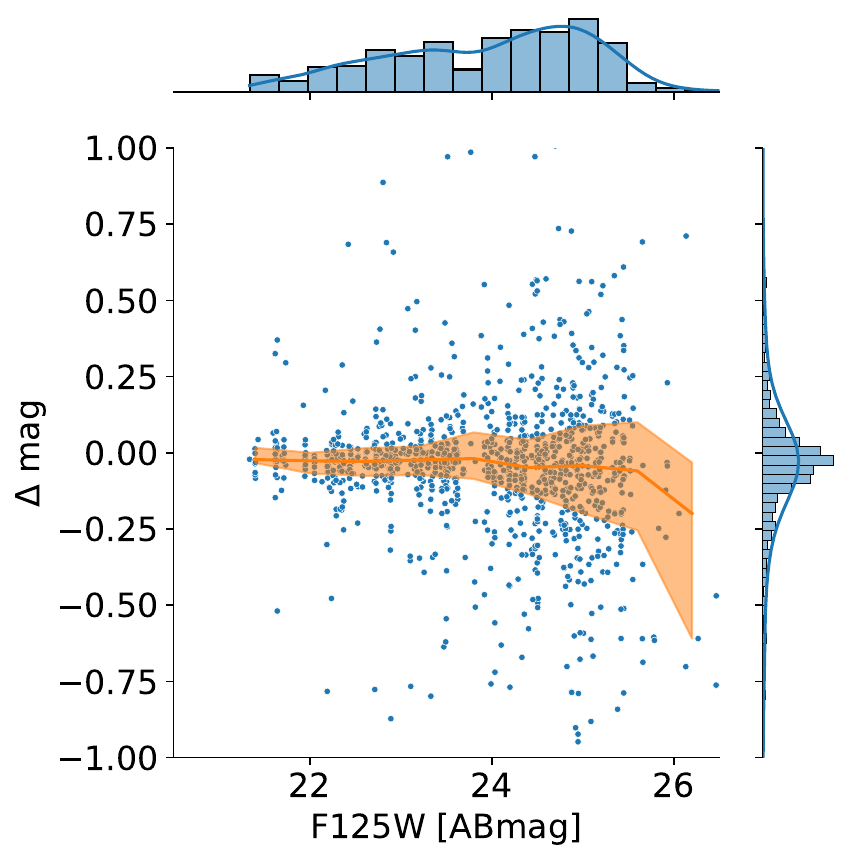}
    \includegraphics[width=0.3\textwidth]{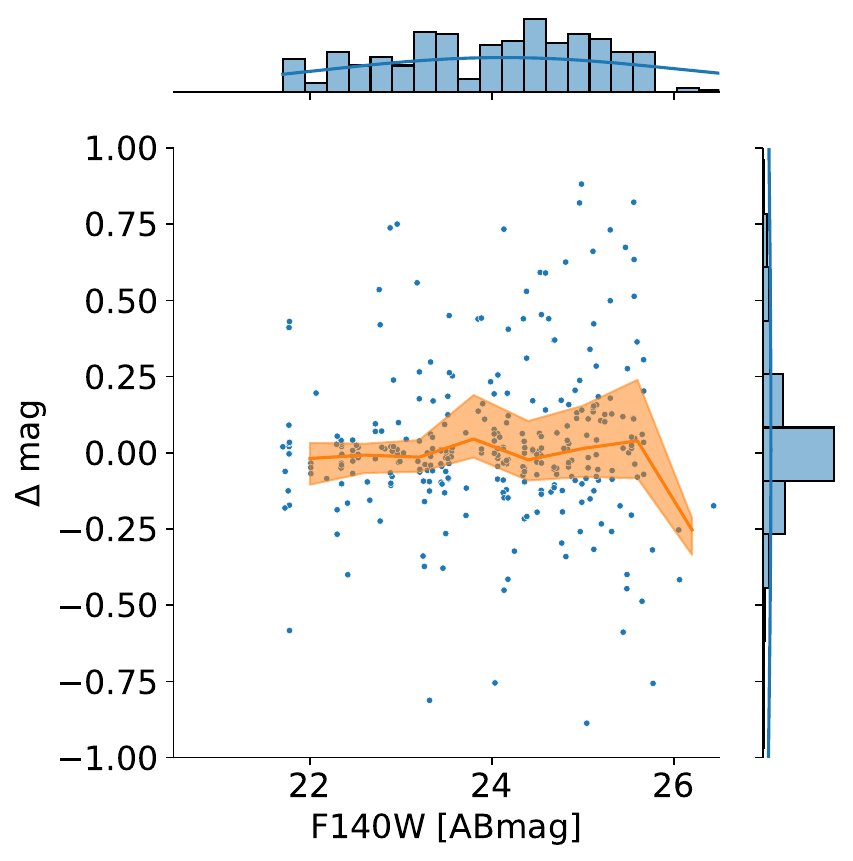}
    \includegraphics[width=0.3\textwidth]{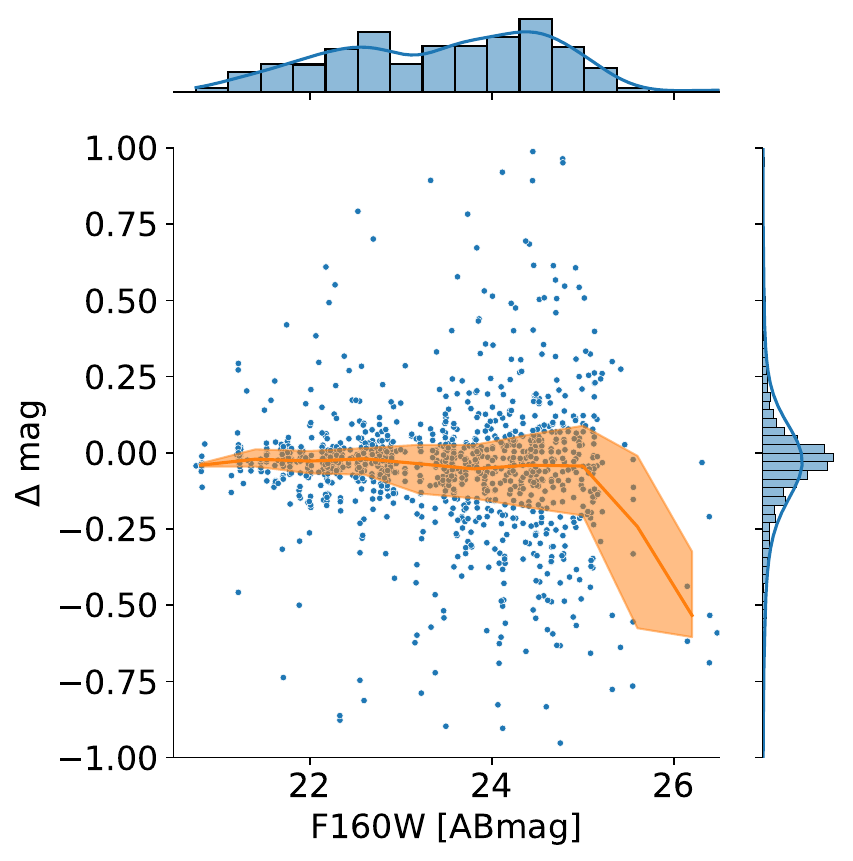}
    \includegraphics[width=0.3\textwidth]{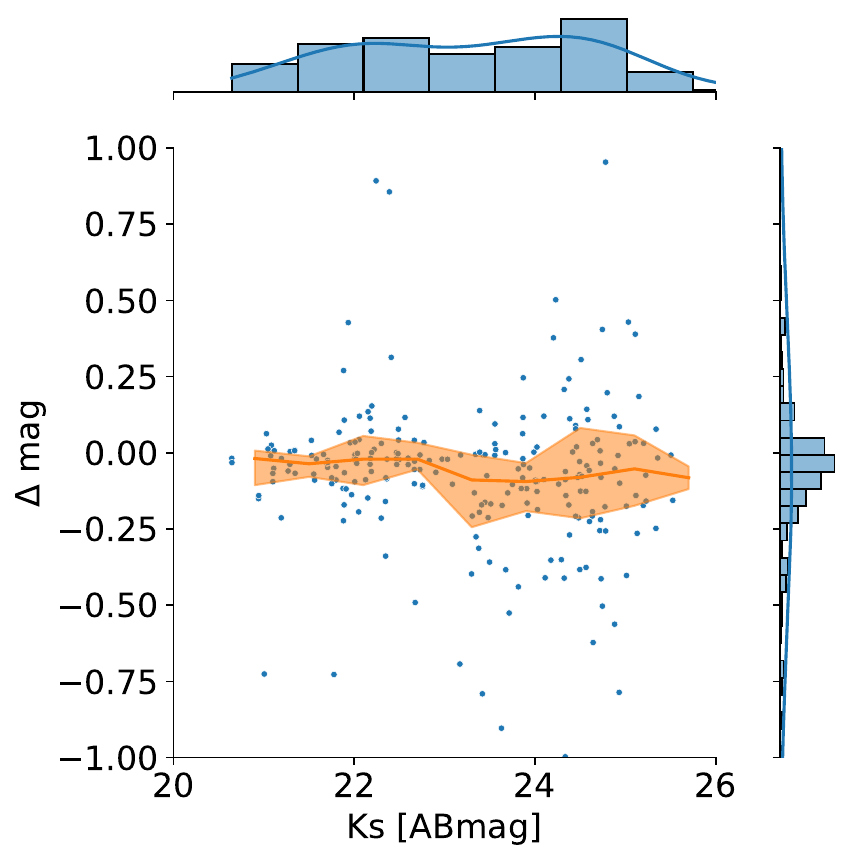}
    \includegraphics[width=0.3\textwidth]{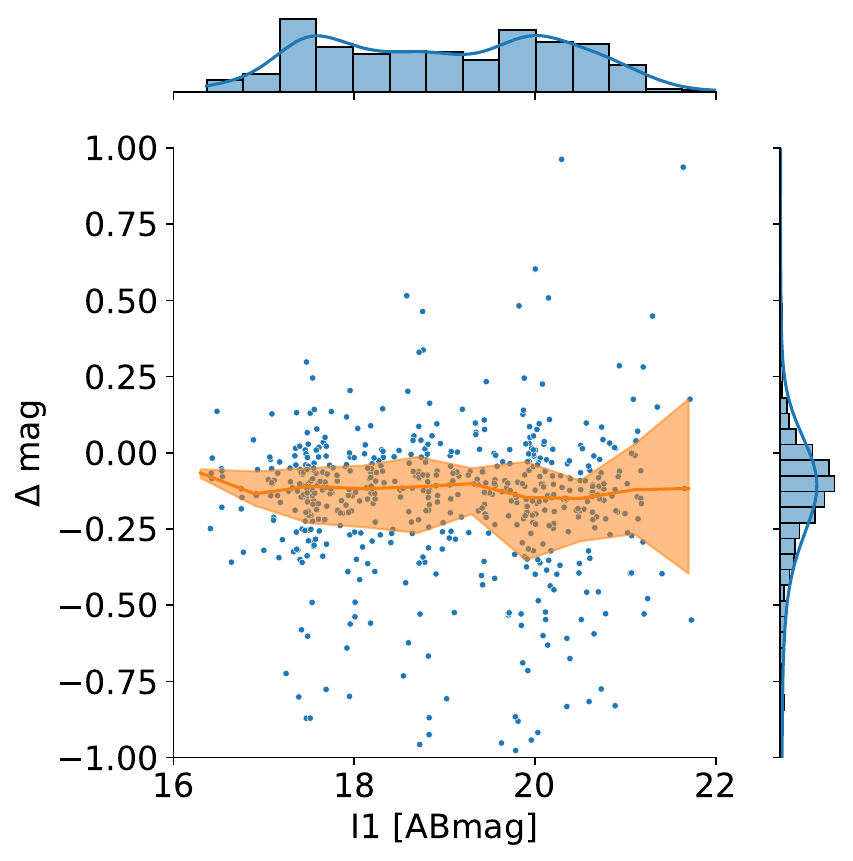}
    \includegraphics[width=0.3\textwidth]{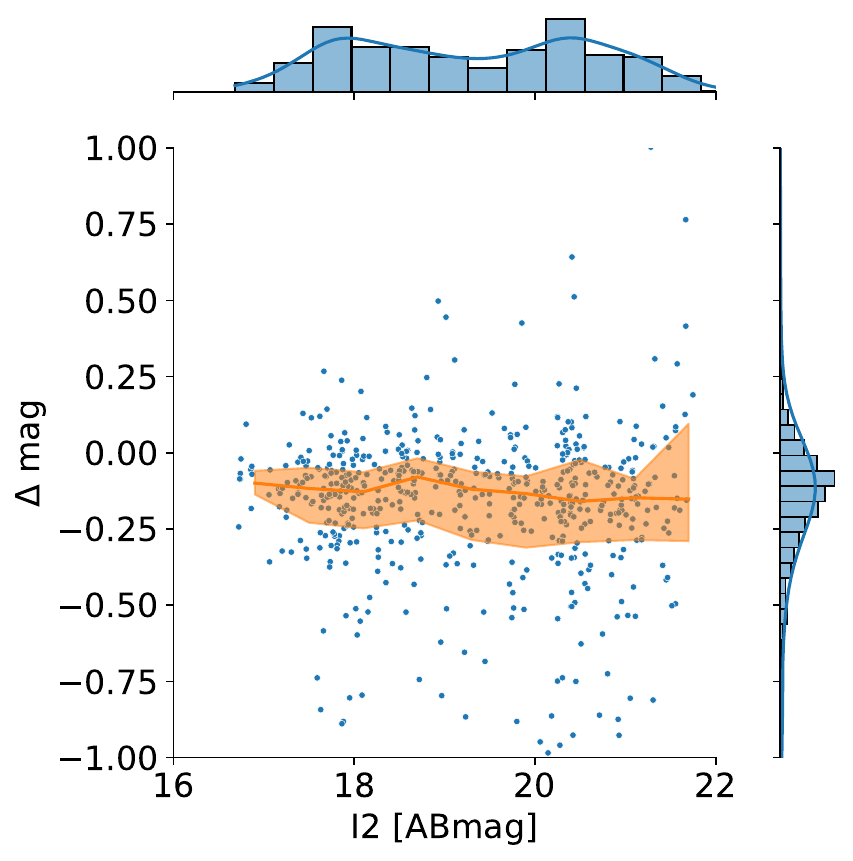}
    \caption{Photometric validation of the BUFFALO catalogs. Scatter plot of $\Delta \rm{mag} = \rm{mag}_{in} - \rm{mag}_{out}$ as a function of the input magnitude for the different bands considered in this work. The solid line shows the rolling mean magnitude offset and the shaded area corresponds to the inter-quartile range. Additionally, each panel includes the input magnitude histogram (top horizontal histogram) as well as the $\Delta \rm{mag}$ histogram (right vertical histogram).}
    \label{fig:photo_validation}
\end{figure*}

This provides valuable information about completeness and absolute zeropoint calibration.  The two catalogs are matched using a nearest neighbor matching routine, \texttt{match\_coordinates\_sky}, included in the \texttt{astropy} package~\citet{astropy:2013, astropy:2018}.  The results of this comparison are shown in Figure~\ref{fig:photo_validation}. We see that for all of the \textit{HST} bands (F435W, F606W, F814W, F105W, F125W, F140W, F160W) the recovered magnitude is within 20 mmags of the input, and that the reconstruction of the fluxes is relatively stable across the considered range of magnitudes.  We note that at the bright end, there is a small fraction of the flux missing, probably due to the extended tails of the sources not being captured by the aperture. This photometric bias becomes smaller with increasing magnitude up to the point where we start to lose sensitivity. We use these offsets to robustly correct the fluxes in each band. For Ks the performance is also excellent and we find a median value of $\Delta \rm{mag}=-0.05 \rm{mag}$. For the Spitzer IRAC channels, we find a small photometric offset $\Delta \rm{mag} = -0.12$ and $\Delta \rm{mag} = -0.13$ for I1 and I2, respectively. 

We compare the mean uncertainty reported by the measurement pipeline to the standard deviation of $\Delta \rm{mag}$ as a function of magnitude. Again, for the \textit{HST} bands the performance is excellent, and we find that the reported errors are in good agreement with the scatter measured using our synthetic sources. This is not the case for Ks nor IRAC, where we find that a correction is needed. In particular, we use a power-law correction:
\begin{equation}
    \Delta F_{new} = \Delta F_{old} AF^{B},
\end{equation}
where $\Delta F_{new}$ is the corrected uncertainty estimate, $\Delta F_{old}$ is the reported uncertainty by the measurement software, $F$ is the reported flux, and $A$, $B$ are free parameters. We fit $A$, $B$ and tabulate the results in Table~\ref{tab:fit_coeffs}.

\begin{table}[h]
    \centering
    \begin{tabular}{c|c|c}
       Band  & $A$ [counts/s]$^{-1}$ & $B$  \\
       \hline
        Ks & 2.05 & 0.26\\
        I1 & 164.67 & 0.44\\
        I2 & 123.14 & 0.43\\
    \end{tabular}
    \caption{Best-fit coefficients used to perform the uncertainty correction as a function of flux.}
    \label{tab:fit_coeffs}
\end{table}

\section{Data products and results}
In this section, we discuss the data products from this work and present some validation results. We produce several new data products from BUFFALO, including catalogs, models for the point spread function, and models for the ICL and bright galaxies. The final catalogs include properties of >100,000 sources in the 6 BUFFALO cluster and parallel fields, and extend the Frontier Fields footprint, covering a total of $\sim240$ square-arcminutes. These include positions, multi-waveband photometry, and photometric redshift estimates for the sources detected as provided by \textsc{LePhare}~\citep{LePhare1999, LePhare2006}.  
Additional details about the information provided by these catalogs can be found in Appendix~\ref{app:columns}.

\begin{figure}
    \centering
    \includegraphics[width=0.9\columnwidth]{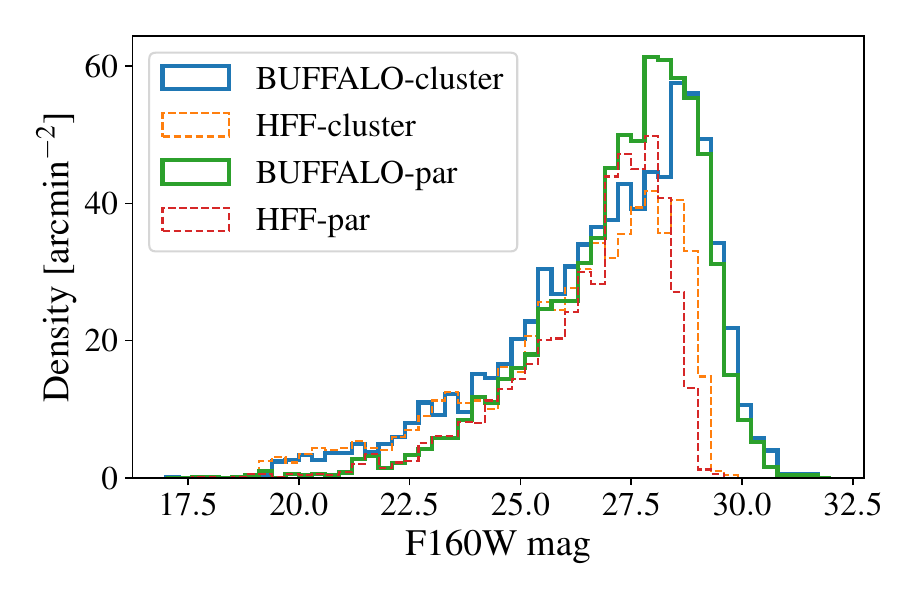}
    \includegraphics[width=0.9\columnwidth]{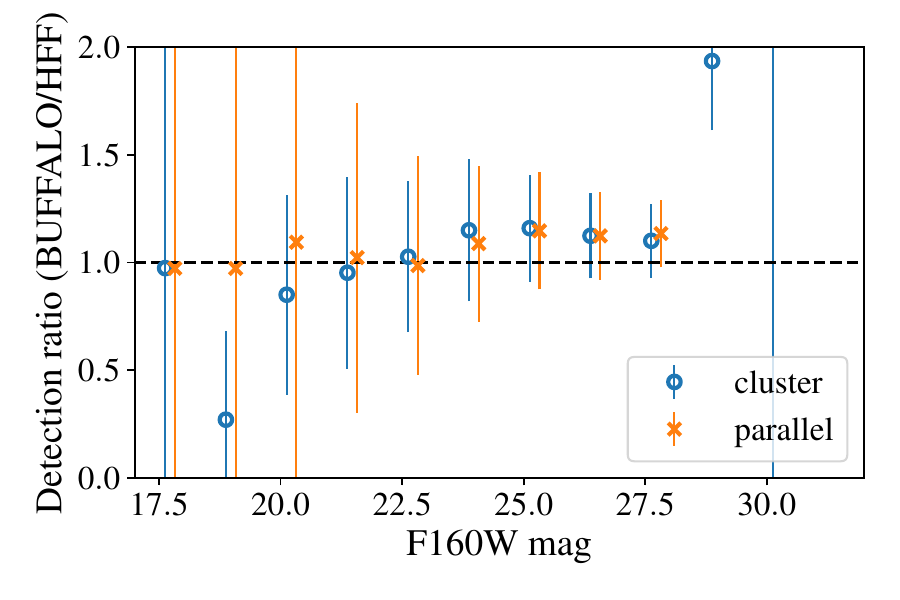}
    \caption{\textbf{Top panel:} Comparison of magnitude distribution in the F160W band between BUFFALO sources (solid lines) and HFF (broken lines) for both the cluster and parallel (par) fields in the deepest part of the images for MACS J1149. \textbf{Bottom panel:} Ratio between the histograms in the top panel for the cluster field (blue open circles) and parallel field (orange crosses). Other fields show similar behavior}
    \label{fig:comp_magdist}
\end{figure}


Point spread function (PSF) estimates are provided as as FITS images. Section~\ref{ssec:psf} describes the modeling of the PSFs. We summarize some of their properties in Table~\ref{tab:psffwhm}. Unsurprisingly, these results are very similar to those found by P21, as the BUFFALO fields are mostly extensions of the HFF.

The procedure to obtain models for the ICL and bright galaxies is described in Section~\ref{ssec:bcg_icl_model}. These models are also available as FITS images.

\subsection{Photometric redshifts}
\label{sec:photz}

In this section we present our redshift estimates based on the photometric measurements presented in previous sections. We run \textsc{LePhare} ~\citep{LePhare1999, LePhare2006}, a template-based code that derives a redshift likelihood function for each source.  As in P21, the fluxes used as inputs to \textsc{LePhare} are rescaled by a factor:
\begin{equation}
    f_\mathrm{tot} =  \frac{\sum_i w_i (\mathrm{FLUX\_AUTO}/\mathrm{FLUX\_ISO})_i}{\sum_i w_i},
\end{equation}
i.e.\ the weighted mean of the AUTO-to-ISO flux ratio summed over the observed \textit{HST} bands, where the weights, $w_{i}$, are the sum in quadrature of the \textsc{Source Extractor} errors: $w_{i}= \sqrt{\sigma_{i,\mathrm{AUTO}}^2 +\sigma_{i,\mathrm{ISO}}^2}$. This is done in order to improve the accuracy of the colors. For the \texttt{TPHOT}-based photometry (Ks, and IRAC bands), as we do not have an equivalent to FLUX\_ISO, we include our baseline fluxes. The template library, and dust attenuation follows~\cite{laigle16}, using \citet{prevot84} or \citet{calzetti2000} extinction laws depending on the galaxy type. For details about the templates and the extinction prescriptions we refer the reader to~\cite{laigle16} and P21. In our catalog the redshift estimates, \texttt{ZPDF}, correspond to the position of the maximum-likelihood for each object.

The redshift calibration procedure is similar to that presented in P21, which is based on spectroscopic data described in \citet[][]{owers11,ebeling14,richard14,balestra16,treu15,schmidt14,2015MNRAS.452.1437J,grillo16,treu16,lagattuta17,mahler18,lagattuta19}. We obtain the best-fit template for each source and try to find a systematic offset in each band by comparing the predicted and observed flux for all sources that have a measured spectroscopic redshift with a spectroscopic quality flag $>3$. These magnitude offsets, when applied to the photometric baseline, compensate for a possible bias in the template library and/or for calibration issues in data reduction. We find these corrections to be below 9\% for all the \textit{HST} bands. For the $K_\mathrm{s}$ band, we find a correction of $0.883$ while in the IRAC channels 1 and 2, the correction is a factor $1.117$ and $1.182$, respectively. These corrections are shown in Table~\ref{tab:photo-calib}. 

\begin{figure*}
    \centering
   
    \includegraphics[width=\textwidth]{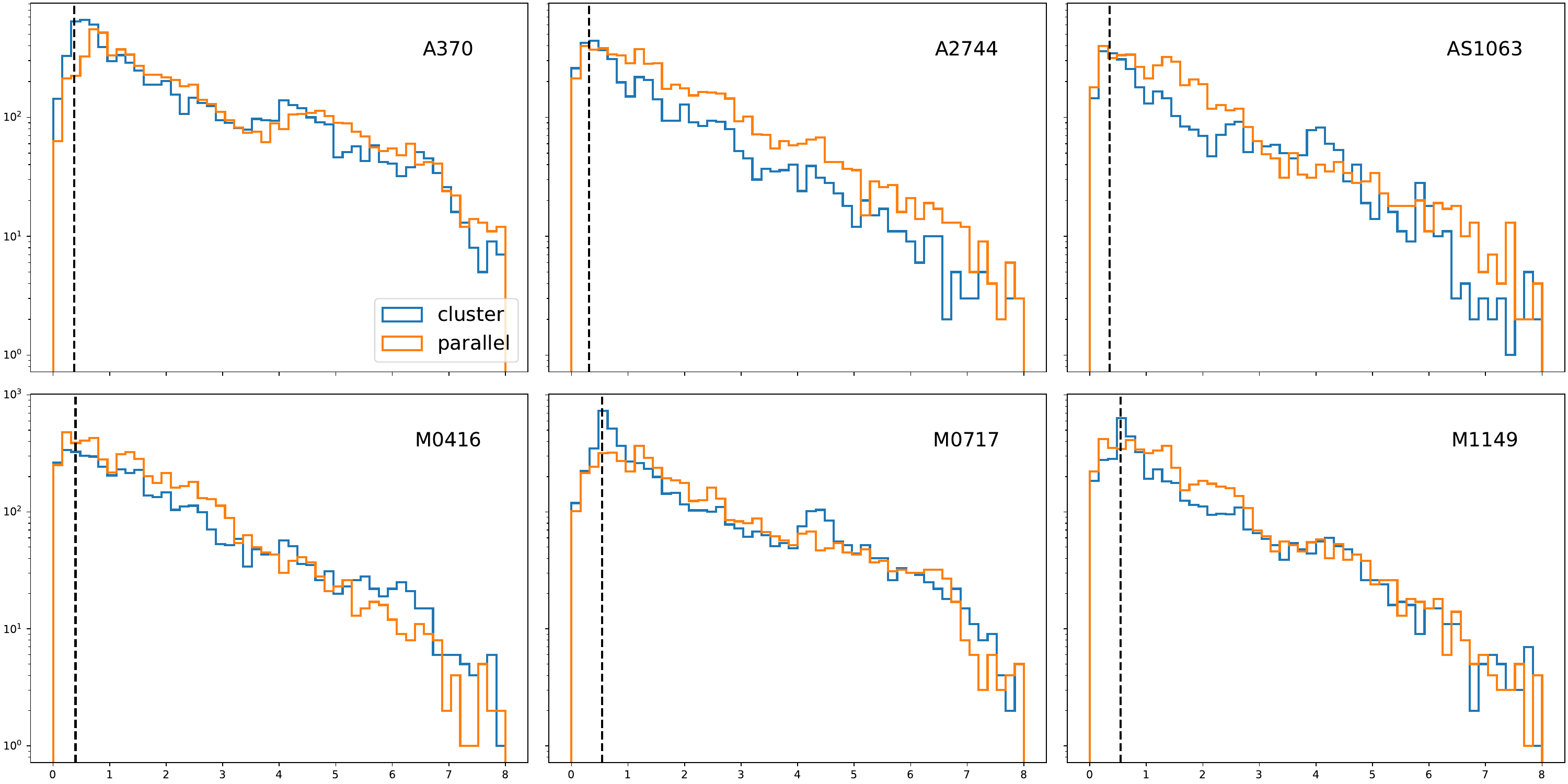}

    \caption{Redshift histograms for each cluster field. Dotted line corresponds to the redshift of the cluster. Sources were chosen with mag$AB_{F160W}<28.5$ and $\chi^2/\rm{ndof}<5$.}
    \label{fig:photz_hists}
\end{figure*}

Figure~\ref{fig:photz_hists} also shows the photometric redshift distribution for objects in each cluster, estimated from the SED fits with a reduced $\chi^2<$ 10.

\begin{deluxetable}{lc}[h!]

\tablecaption{Multiplicative factors applied to each band in the photo-z calibration step.}
\label{tab:photo-calib}
\tablehead{\colhead{Band} & \colhead{Multiplicative Factor}\hspace{.5cm}}

\startdata
F275W & 1.055\\
F336W & 1.011\\
F435W & 1.085 \\
F475W & 1.060 \\
F606W & 1.004 \\
F625W & 1.006 \\
F814W & 0.992 \\
F105W & 1.004 \\
F110W & 1.015 \\
F125W & 1.011 \\
F140W & 1.008 \\
F160W & 0.995 \\
Ks & 0.883 \\
IRAC1 & 1.117 \\
IRAC2 & 1.182 \\
\enddata

\end{deluxetable}




\section{Comparison with the Hubble Frontier Fields}
\label{ssec:comparison_hff}
By design, there is significant overlap between the HFF and the BUFFALO fields. This makes the HFF catalogs an exceptional reference to verify and validate the data presented in this work and to check for potential improvements, given the increased number of exposures. Here, we compare our BUFFALO data products with those presented in P21.

Figure~\ref{fig:comp_magdist} compares the magnitude distribution of sources in the F160W band between the catalog presented here and the catalogs in P21 in the overlapping region of the MACS J1149 cluster. Here we show that our new BUFFALO catalogs reach fainter sources than those from the HFF. We also show the fraction of detected objects as a function of magnitude, finding that both catalogs have a similar completeness to magnitude $\sim 27.5$ in the F160W band. This is in agreement with P21, where the completeness dropped below 100\% at $\sim 27.5$. Other bands and clusters show a similar behavior. We note that these completeness estimates do not take into account the effects of strong lensing.




\begin{figure}
    \centering
    \includegraphics[width=0.5\textwidth]{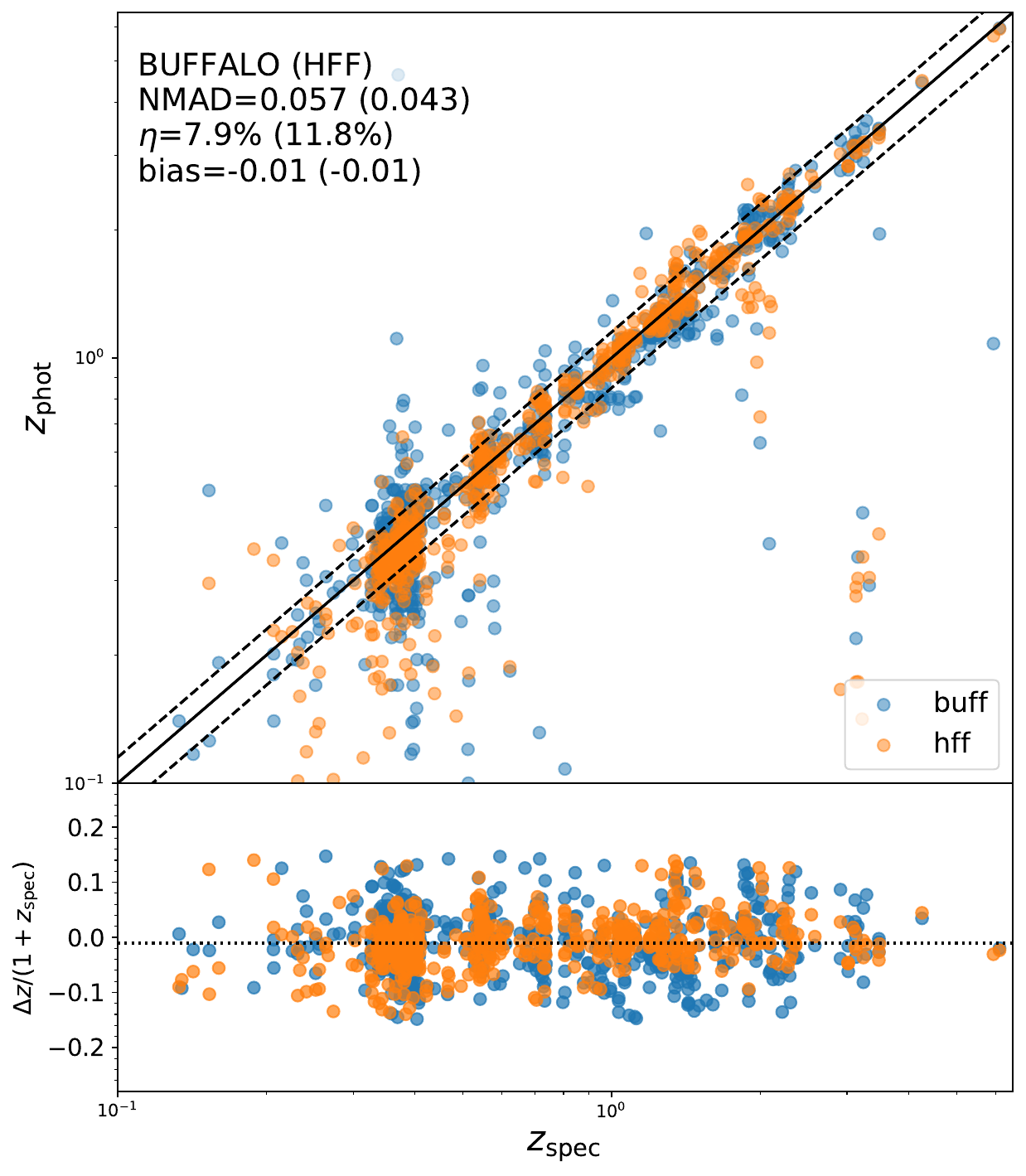}
    \caption{Assessing the quality of photometric redshifts estimated through SED fitting. \textbf{Top panel:} Photometric vs. spectroscopic redshift
comparison. Blue and orange points are 709 matched BUFFALO and HFF sources, respectively, whose with BUFFALO F160W band magnitude are between 16 and 26, are detected in at least 5 BUFFALO bands, with a redshift quality flag > 3. The solid line shows the 1:1 relationship,
and the dashed lines encloses the $z_{phot} = z_{spec} \pm 0.15(1+z_{spec})$
threshold used to identify outliers (i.e., catastrophic errors).
NMAD scatter ($\sigma$) and outlier fraction ($\eta$) are reported on
the top-left corner. \textbf{Bottom panel:} $\Delta z \equiv z_{phot} - z_{spec}$ scatter.}
    \label{fig:photzspecz}
\end{figure}

\begin{figure}
    \centering
    \includegraphics[width=0.5\textwidth]{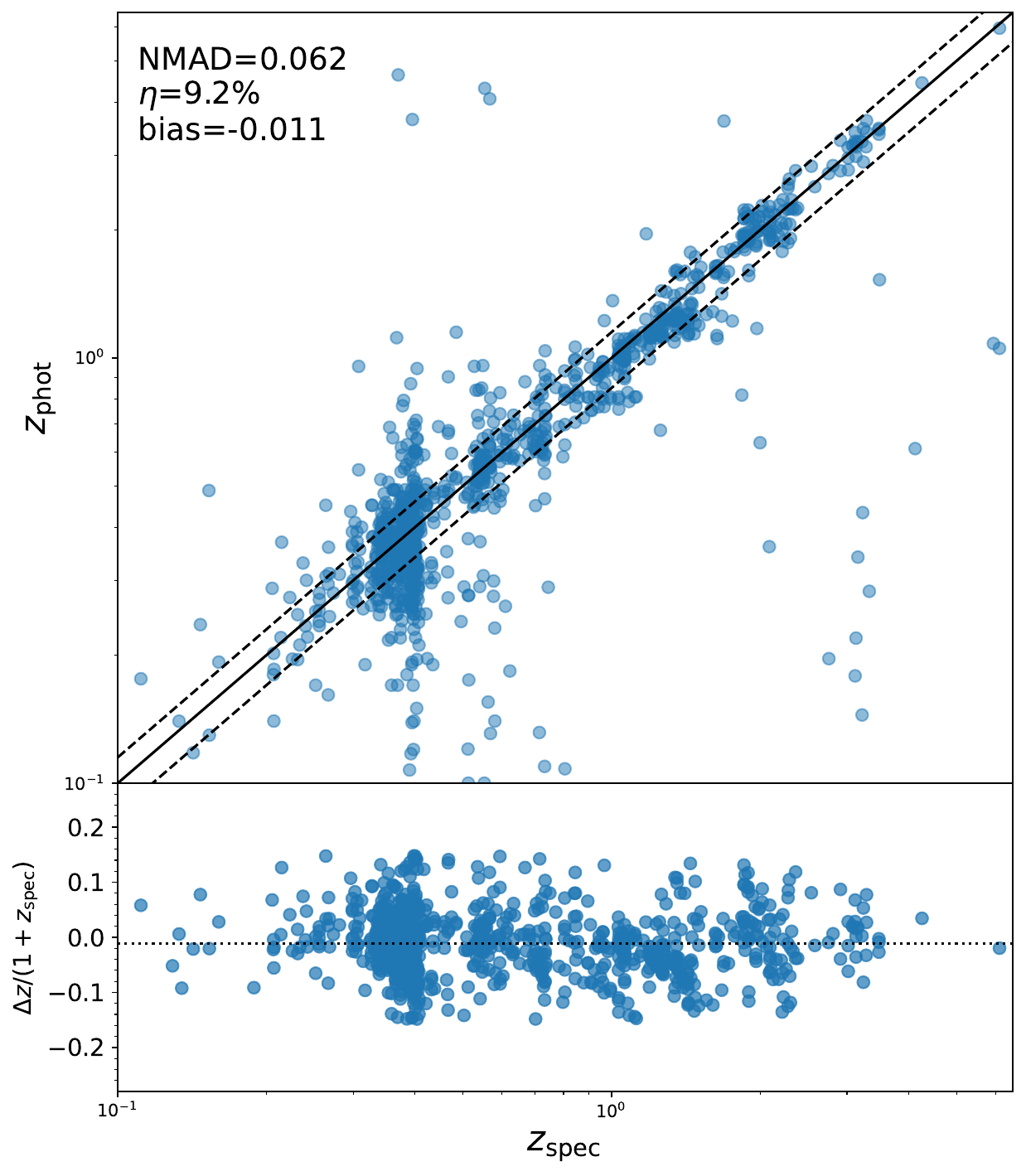}
    \caption{Assessing the quality of photometric redshifts estimated through SED fitting. \textbf{Top panel:} Photometric vs. spectroscopic redshift
comparison. Blue points are 999 spectroscopic redshifts selected for
16 < F 160W < 26, detected in at least 5 bands, with a redshift quality flag > 3. The solid line shows the 1:1 relationship
and the dashed lines encloses the zphot = zspec$\pm$0.15(1$+$zspec)
threshold used to identify outliers (i.e., catastrophic errors).
The median of $\Delta z/(1+z_{spec})$ (bias) and outlier fraction ($\eta$) are reported on
the top-left corner. \textbf{Bottom panel:} $\Delta$z $\equiv$ zphot $-$ zspec scatter.}
    \label{fig:photzspecz}
\end{figure}

\section{Summary}
The wealth of deep (\textit{HST}) observations and ancillary data in the HFF~\citep[][]{2017ApJ...837...97L}, open a window to the high-redshift universe, and provides a complementary sample to the \emph{JWST}. The BUFFALO survey~\citep[PIs: Mathilde Jauzac, Charles Steinhardt]{steinhardt20} used these data and extended the observations in the 6 HFFs, to allow for follow-up spectroscopy. This work presents a new set of data products based on the BUFFALO observations. The data products include models for the point spread function (PSF), intra-cluster light (ICL), the bright galaxies, and catalogs of astronomical sources. The catalogs contain detailed information (including positions and photometry) of over 100,000 sources distributed across 6 separate cluster and parallel fields covering a total area of 240 arcmin$^{2}$. 

The data products are obtained using a similar procedure to that outlined in~\citet{2021ApJS..256...27P}. First, a model of the bright galaxies, and the ICL are created. These models are then subtracted from the original image, in order to increase our sensitivity allowing us to observe fainter sources, which are detected and measured using \textsc{Source Extractor} in the \textit{HST} bands. We then use the IR-weighted segmentation map as priors in the \textsc{T-PHOT} package to obtain forced-photometry in ancillary data from Keck $K$s band, and Spitzer IRAC channels 1 and 2. The photometric measurements are validated using synthetic source injection. Finally, \textsc{LePhare} is run to obtain redshift estimates based on our photometric measurements. The main change with respect to the procedure in P21 is the usage of a ``super hot'' mode Source Extractor run, that simplifies bookkeping, while not biasing the photometric estimates. As a sanity check, we plot the redshift histograms and note that the peaks of these histograms correspond to the redshift of each respective cluster.

This catalog represents one of the deepest views at galaxy clusters to date and a sample that lends itself well for \textit{JWST} follow-up. All of the data products presented in this work will be made publicly available to the astronomical community through the usual astronomical archive databases (MAST and Vizier).

\section*{Acknowledgements}
ID acknowledges the support received from the European Union's Horizon 2020 research and innovation programme under the Marie Sk\l{}odowska-Curie grant agreement No. 896225. This work has made use of the CANDIDE Cluster at the Institut d'Astrophysique de Paris and made possible by grants from the PNCG and the DIM-ACAV. The Cosmic Dawn Center is funded by the Danish National Research Foundation under grant No. 140. LF acknowledges support by Grant No. 2020750 from the United States-Israel Binational Science Foundation (BSF) and Grant No. 2109066 from the United States National Science Foundation (NSF).

Based on observations made with the NASA/ESA \textit{Hubble Space Telescope}, obtained at the Space Telescope Science Institute, which is operated by the Association of Universities for Research in Astronomy, Inc., under NASA contract NAS 5-26555. These observations are associated with programs GO-15117, WFC3/UV imaging (GO 13389, 14209; B. Siana), A370 HST/ACS additional imaging (GO 11507; K. Noll, 11582; A. Blain, 13790; S. Rodney, 11591; J.P. Kneib)

This work is based in part on data and catalog products from HFF-DeepSpace, funded by the National Science Foundation and Space Telescope Science Institute (operated by the Association of Universities for Research in Astronomy, Inc., under NASA contract NAS5-26555).

Support for \textit{HST} Program GO-15117 was provided through a grant from the STScI under NASA contract NAS5-26555.

This work is based in part on observations made with the Spitzer Space Telescope, which was operated by the Jet Propulsion Laboratory, California Institute of Technology under a contract with NASA.

Based on observations collected at the European Organisation for Astronomical Research in the Southern Hemisphere under ESO programme(s) 090.A-0458, 092.A-0472,
and 095.A-0533.

Some of the data presented herein were obtained at the W. M. Keck Observatory, which is operated as a scientific partnership among the California Institute of Technology, the University of California and the National Aeronautics and Space Administration. The Observatory was made possible by the generous financial support of the W. M. Keck Foundation.

The authors wish to recognize and acknowledge the very significant cultural role and reverence that the summit of Maunakea has always had within the indigenous Hawaiian community.  We are most fortunate to have the opportunity to conduct observations from this mountain.

\bibliography{bib}

\appendix
\section{Catalog details}
\label{app:columns}
The catalogs presented in this work contain the following information:
\begin{itemize}
    \item ID: Source number 
    \item FLUX\_FXXXW: Total scaled flux in cgs units of $\mathrm{erg/cm}^2\mathrm{/s/Hz}$ 
    \item FLUXERR\_FXXXW: Corrected flux error in cgs units of $\mathrm{erg/cm}^2\mathrm{/s/Hz}$ 
    \item ZSPEC: reported spectroscopic redshift
    \item ZSPEC\_Q: reported quality flag of spectroscopic redshift
    \item ZSPEC\_REF: dataset from which spectroscopic redshift was obtained
    \item ALPHA\_J2000\_STACK: Right Ascension (J2000) in degrees using GAIA DR2 as reference.
    \item DELTA\_J2000\_STACK: Declination (J2000) in degrees using GAIA DR2 as reference.
    \item FIELD: denotes the field object belongs to
    \item ZCHI2: photometric redshift goodness of fit
    \item CHI2\_RED: reduced chi square
    \item ZPDF: photometric redshift derived via maximum likelihood 
    \item ZPDF\_LOW: lower threshold for photometric redshift
    \item ZPDF\_HIGH: upper threshold for photometric redshift
    \item MOD\_BEST: galaxy model for best $\chi^2$ 
    \item EXT\_LAW: Extinction law  
    \item E\_BV: E(B-V)
    \item ZSECOND: secondary photometric redshift peak in maximum likelihood distribution
    \item BITMASK: Base 2 number to determine which bands were used. Calculated via bitmask$=\sum_{n=good band index} 2^n$
    \item NB\_USED: number of bands used
\end{itemize}

\section{Source Extractor Configuration}
\label{app:Source Extractor_config}
\texttt{DETECT\_MINAREA}  3

\texttt{DETECT\_THRESH}   0.5

\texttt{ANALYSIS\_THRESH} 0.5

\texttt{FILTER}         Y 

\texttt{FILTER\_NAME}     gauss\_4.0\_7x7.conv 

\texttt{DEBLEND\_NTHRESH} 64 

\texttt{DEBLEND\_MINCONT} 0.000005 

\texttt{CLEAN}          Y     

\texttt{CLEAN\_PARAM}     0.8 

\texttt{MASK\_TYPE}       CORRECT         

\texttt{PHOT\_AUTOPARAMS} 2.0, 3.5 

\texttt{PHOT\_FLUXFRAC}   0.5

\texttt{SEEING\_FWHM}     0.17 

\texttt{STARNNW\_NAME}    goods\_default.nnw 

\texttt{BACK\_SIZE}       64

\texttt{BACK\_FILTERSIZE} 3

\texttt{BACKPHOTO\_TYPE}  LOCAL

\texttt{BACKPHOTO\_THICK} 24 

\end{document}